\documentclass[12pt]{article}
\pdfoutput=1
\usepackage{amsmath, amsthm,comment}
\usepackage{amsfonts}
\usepackage{amssymb,graphics,psfrag}
\usepackage{array,epsfig,stmaryrd,graphicx}
\usepackage{cite}
\usepackage{graphicx}
\usepackage{makeidx}
\usepackage{multicol}
\usepackage{geometry}
\usepackage{amsfonts}
\usepackage{mathrsfs}
\usepackage{amssymb}
\usepackage{amsmath}
\usepackage{slashed}
\usepackage{cancel}
\usepackage{heck}%
\providecommand{\U}[1]{\protect\rule{.1in}{.1in}}
\newcommand{\beq}{\begin{equation}}
\newcommand{\eeq}{\end{equation}}
\newcommand{\be}{\begin{equation}}
\newcommand{\ee}{\end{equation}}
\newcommand{\bea}{\begin{eqnarray}}
\newcommand{\eea}{\end{eqnarray}}
\newcommand{\ben}{\begin{eqnarray*}}
\newcommand{\een}{\end{eqnarray*}}
\newcommand{\ba}{\begin{aligned}}
\newcommand{\ea}{\end{aligned}}
\newcommand{\bt}{\begin{tabular}}
\newcommand{\et}{\end{tabular}}
\newcommand{\bc}{\begin{center}}
\newcommand{\ec}{\end{center}}

\newcommand{\cref}{{\bf [check ref]}}

\newcommand{\bs}{\begin{subarray}{c}}
\newcommand{\es}{\end{subarray}}

\begin{document}

\date{June, 2010}
\title{An Exceptional Sector for F-theory GUTs}

%\preprint{arXiv:????.????}

\institution{IAS}{\centerline{${}^{1}$School of Natural Sciences, Institute for Advanced Study, Princeton, NJ 08540, USA}}

\institution{HarvardU}{\centerline{${}^{2}$Jefferson Physical
Laboratory, Harvard University, Cambridge, MA 02138, USA}}

\authors{Jonathan J. Heckman\worksat{\IAS}%
\footnote{e-mail: \texttt{jheckman@sns.ias.edu}%
} and Cumrun Vafa\worksat{\HarvardU}%
\footnote{e-mail: \texttt{vafa@physics.harvard.edu}%
}}

\abstract{D3-branes are often a necessary ingredient in global
compactifications of F-theory.  In minimal realizations of flavor hierarchies
in F-theory GUT models, suitable fluxes are turned on,
which in turn attract D3-branes to the Yukawa points. Of particular importance are ``E-type'' Yukawa points,
as they are required to realize a large top quark mass. In this paper we study the worldvolume theory of a
D3-brane probing such an E-point. D3-brane probes of isolated exceptional singularities
lead to strongly coupled $\mathcal{N} = 2$ CFTs of the type found by Minahan and Nemeschansky.
We show that the local data of an E-point probe theory determines an
$\mathcal{N} = 1$ deformation of the original $\mathcal{N} = 2$ theory which couples this strongly interacting
CFT to a free hypermultiplet.  Monodromy in the seven-brane configuration translates to a novel
class of deformations of the CFT.
We study how the probe theory couples to the
Standard Model, determining the most relevant F-term couplings, the effect of the probe on the running
of the Standard Model gauge couplings, as well as possible sources of kinetic mixing with the Standard Model.}

%TCIMACRO{\TeXButton{Maketitle}{\maketitle}}%
%BeginExpansion
\maketitle
%EndExpansion

\enlargethispage{\baselineskip}
\tableofcontents

\section{Introduction}

The existence of the landscape is a significant impediment to bridging the gulf
between strings and particle phenomenology. One way to narrow the search
for promising, predictive vacua is to demand that the visible sector
admits the structures of a Grand Unified Theory (GUT), but also remains
decoupled from gravity. Imposing both conditions turns out to be quite stringent,
but remarkably, can be satisfied in local F-theory GUT
models \cite{BHVI,BHVII}.

In F-theory GUTs (see \cite{DWI,BHVI,WatariTATARHETF,BHVII,DWII} and the references in the review
\cite{Heckman:2010bq} for a partial list), the visible gauge sector arises on the worldvolume of a seven-brane
wrapping a four-manifold $S$ of positive curvature, with inverse radius set by the GUT scale.
The intersection of the gauge seven-brane with additional seven-branes can
lead to matter fields living on curves inside $S$. The intersection of matter curves
can lead to Yukawa couplings localized on points in $S$.

Decoupling gravity from this system imposes strong restrictions on the low energy content of
such models \cite{DWI,DWII,Andreas:2009uf,DWIII,Cordova:2009fg,Grimm:2009yu,HeckVerlinde}.
On the other hand, a natural expectation from low energy theory field theories
is that at higher energies, the Standard Model could couple to other sectors, with potentially
interesting consequences for phenomenology. Recent examples include coupling
the Standard Model to hidden valleys \cite{Strassler:2006im}, or even an ``unparticle''
sector \cite{Georgi:2007ek}.

This raises the question: What additional sectors can there be once we have
decoupled an F-theory GUT from the bulk dynamics of a string compactification?
As emphasized in \cite{EPOINT}, seven-branes from E-type structures are naturally shielded from additional
seven-brane sectors, as there is ``nowhere to go'' beyond $E_8$. Probe D3-branes, however, naturally provide
an additional sector since their presence is often required in order to cancel tadpoles in global
models \cite{Sethi:1996es}.

In this paper we study the additional sector provided by a probe D3-brane.
The worldvolume theory on a D3-brane is captured by the choice of where it sits
in the internal geometry of the F-theory threefold base $B$. At a generic point of $B$,
this leads to a free $U(1)$ theory with ${\cal N}=4$ supersymmetry.
This is not very interesting because it leads to a trivial free theory in the infrared (IR), and
is also decoupled from the visible sector, as a generic point is far away
from the Standard Model seven-branes.

Considerations from flavor physics substantially modify this ``generic'' picture.
As found in \cite{HVCKM}, fluxes induce hierarchical corrections to the leading order rank one
structure of Yukawa couplings, providing a natural mechanism for generating flavor hierarchies \cite{HVCKM}.
A detailed study of this proposal revealed that the requisite fluxes also induce a superpotential
for D3-branes which \textit{attracts} the D3-brane to a Yukawa point \cite{FGUTSNC}.

In order to realize a large top quark mass, it is necessary to include an E-type point of enhancement.
At the very least, this requires enhancement to an $E_6$ singularity \cite{BHVI} (see also \cite{Tatar:2006dc}).
Moreover, considerations from flavor physics suggest a common origin for \textit{all} Yukawas from a single point
of $E_8$ \cite{BHSV,EPOINT} (see also \cite{HVCKM}). Combining this with the expectation
that D3-branes are attracted to Yukawa points, we are thus led to study the worldvolume
theory of a D3-brane probing an E-type singularity.

D3-brane probes of F-theory have been studied in various contexts, for example in \cite{Banks:1996nj,MNI,MNII}.
In compactifications of F-theory to eight dimensions, the worldvolume theory of a D3-brane probe of an
E-type singularity realizes the strongly coupled $\mathcal{N} = 2$ superconformal field theories (SCFTs) studied by
Minahan and Nemeschansky \cite{MNI,MNII}. The Minahan-Nemeschansky theories are characterized by an E-type
flavor symmetry group, and a one-dimensional Coulomb branch, parameterizing motion of the D3-brane normal
to the seven-brane. This theory can also be realized via compactification of a six-dimensional theory of tensionless
E-strings \cite{Seiberg:1996vs,Klemm:1996hh,Minahan:1998vr}, as well as via
the worldvolume theory of a zero size instanton of the internal four-dimensional gauge theory of an
exceptional seven-brane \cite{WittenSmall}.

More precisely, the D3-brane probe theory is described by two decoupled systems: One is the strongly coupled
CFT of Minahan-Nemeschansky type, and the other is a free hypermultiplet given by two ${\cal N}=1$ chiral
multiplets $Z_1$ and $Z_2$ which parameterize the position of the D3-brane
parallel to the seven-brane. In this theory, all of the operators organize according to
representations of the flavor symmetry group. From the perspective of an F-theory compactification,
the states of the theory correspond to $(p,q)$ strings and their junctions, indicating
the presence of light electric and magnetic states.

\begin{figure}
[ptb]
\begin{center}
\includegraphics[
height=2.13in,
width=2.917in
]%
{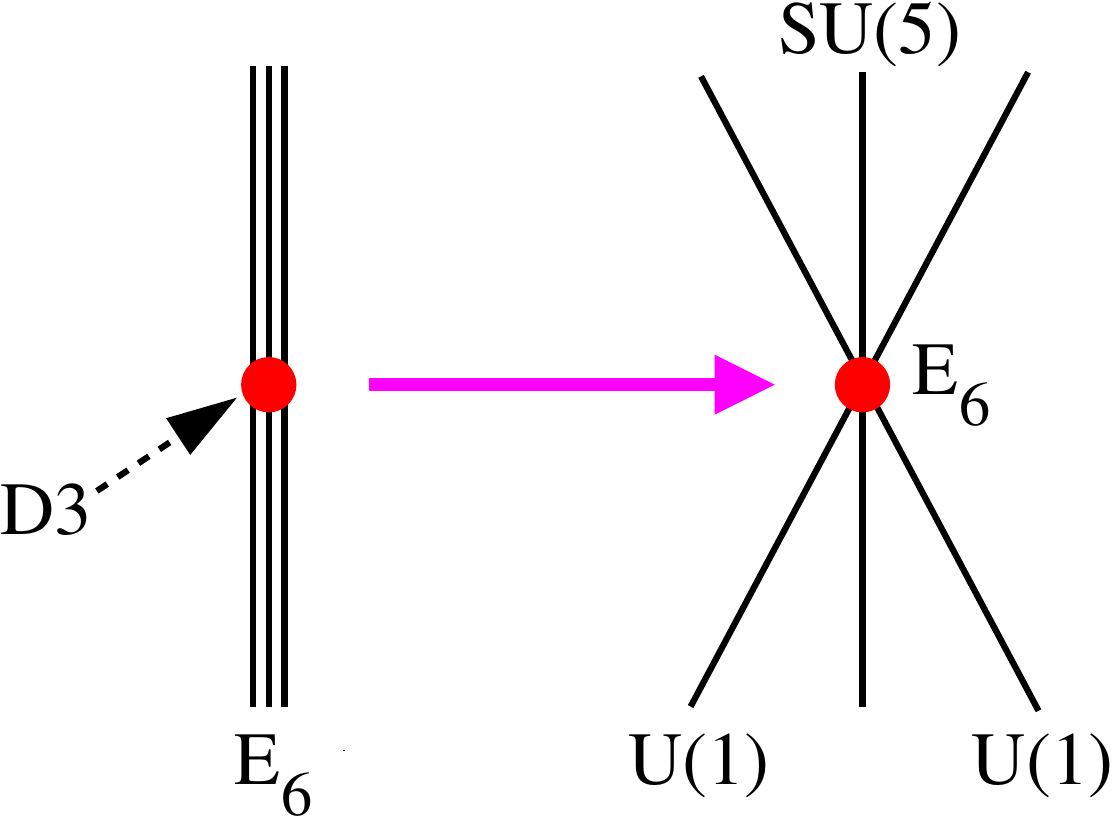}%
\caption{Depiction of a D3-brane probe of a seven-brane with $E_{6}$ gauge
group. In the configuration on the left, we have the Minahan-Nemeschansky
$\mathcal{N}=2$ theory with $E_{6}$ flavor symmetry and a decoupled
$\mathcal{N}=2$ hypermultiplet. A Yukawa point is locally described by tilting
the seven-branes so that they still enhance to $E_{6}$ at the point probed by
the D3-brane. In the probe theory this corresponds to an $\mathcal{N}=1$
deformation which couples the Minahan Nemeschansky theory and free
hypermultiplet.}%
\label{etilt}%
\end{center}
\end{figure}

Since the D3-brane only probes a point of the internal geometry, it is natural to expect some similarities
between the original $\mathcal{N} = 2$ theory, and D3-brane probes of a Yukawa point. Indeed, we can realize the local
behavior of a Yukawa point by tilting a stack of parallel seven-branes in distinct directions of the
threefold base. The probe theory of the Yukawa point corresponds to a D3-brane sitting at
the mutual intersection of these seven-branes. See figure \ref{etilt} for a depiction
of this type of geometry. Translating the geometric and flux data of the compactification
to the probe theory, this corresponds to coupling the two previously decoupled
systems of the ${\cal N}=2$ probe theory, by promoting mass parameters
$m$ transforming in the adjoint representation of the flavor symmetry group $G$
to field dependent operators $m(Z_1,Z_2)$ which depend
on the position of the D3-brane in directions parallel to the GUT seven-brane. In the original
D3-brane probe theory, there are dimension two operators $O$ transforming in the
adjoint representation of $G$. The mass deformation then corresponds to the deformation:
\begin{equation}\label{massdef}
\delta L = \int d^{2} \theta \,\, Tr_{G}(m(Z_1 , Z_2) \cdot O) + h.c..
\end{equation}
In geometric terms, the data defining the deformation is specified along the Coulomb
branch of the probe theory by the Casimirs of $m(\langle Z_1 \rangle , \langle Z_2 \rangle )$, and a
background flux through the seven-branes. In general, we expect that the eigenvalues of
$m(\langle Z_1 \rangle , \langle Z_2 \rangle )$ will
have branch cuts in the dependence on the $z_{i} = \langle Z_i \rangle$.
This is a general phenomenon known as ``seven-brane monodromy''
and leads to a rich class of possible $\mathcal{N}  = 1$ deformations.\footnote{The
analysis we present here also applies to heterotic M-theory compactified on
an interval times an elliptically fibered Calabi-Yau three-fold. Indeed, wrapping a spacetime filling M5-brane along
the elliptic curve corresponds on the F-theory side to a spacetime filling D3-brane. Placing the M5-brane at a point of the complex twofold
base of the Calabi-Yau threefold corresponds to placing a D3-brane at the same point. The unfolding of the singularity
type on the F-theory side in tandem with the flux data from seven-branes
translates to bundle data on the heterotic side. Here we see that when the M5-brane
sits at the analogue of the Yukawa point on the heterotic side, we obtain the same four-dimensional theory engineered on
the F-theory side.}

From the perspective of the seven-brane, the field dependent mass parameter $m(Z_1 , Z_2)$ corresponds to the background value
$\phi_{0}(Z_1 , Z_2)$ of a field $\phi(Z_1 , Z_2)$ which transforms in the adjoint representation of the seven-brane gauge group $G$.
Decomposing $\phi$ into a background and fluctuation as:
\begin{equation}
\phi = \phi_{0} + \delta \phi
\end{equation}
the fluctuations $\delta \phi(Z_1 , Z_2)$ corresponds to matter fields of the visible sector.\footnote{As we
will later explain it is always possible to choose a gauge where all the matter fields are represented by holomorphic
fluctuations of $\phi$.}
The analogue of the mass deformation of equation (\ref{massdef}) is then the F-term coupling:
\begin{equation}
\delta L = \int d^{2} \theta \,\, Tr_{G} ((\phi_{0}(Z_1 , Z_2) + \delta \phi(Z_1 , Z_2) )\cdot O) + h.c..
\end{equation}

If the unfolding leads to an unbroken E-type symmetry, we have an ${\cal N}=1$
field theory with an E-type flavor symmetry. It is rather implausible that a theory with E-type flavor symmetry
flows to an infrared free theory, so it is natural to postulate that at least in these cases,
the infrared limit of these systems flows to an interacting ${\cal N}=1$ CFT. Though we cannot prove it, this suggests that
in all cases where the D3-brane probes an E-point singularity, the IR limit is also an interesting interacting theory. In the
case of trivial monodromy, we can apply the recent analysis of UV marginal deformations given in
\cite{Green:2010da} to prove this is the case. In the more general case, the presence of both electric and magnetic states
(as dictated by the E-type structure) hints that this also holds for more general geometries.

Assuming that we do realize a CFT, it is then natural to ask how this sector couples to the Standard Model. Much as
in the counting of BPS states in the $\mathcal{N} = 2$ case performed for example in \cite{Klemm:1996hh,Gadde:2010te}, there will be an entire tower of
states of different masses and spins which couple to the Standard Model.
The number of such states which couple to the Standard Model
depends on the scale of conformal symmetry breaking $M_{\cancel{CFT}}$.
One can envision many possible applications for the presence of such a strongly coupled CFT, with the specific
application possibly depending on the scale of conformal symmetry breaking. Because we have states
with the gauge quantum numbers of the Standard Model, it follows that we must require $M_{\cancel{CFT}}$ to be greater than
at least a few hundred GeV to avoid conflict with experiment. Modulo this restriction, however,
$M_{\cancel{CFT}}$ could be anywhere from a few hundred GeV up to the GUT scale, or higher. At low energies,
at most only a few particle states of this sector would be observable at the LHC, though the strong
coupling to the D3-brane probe sector might still produce novel signatures. Let us also
note that because these states admit a particle interpretation, this type of scenario is
somewhat different from the unparticle scenario proposed in \cite{Georgi:2007ek}. At intermediate scales,
the D3-brane probe could provide a source of supersymmetry breaking and/or perhaps
a novel messenger sector of a gauge mediation scenario.

Here we deduce the general form of F-term couplings to the Standard Model. In addition,
we determine the effect of this nearly conformal sector on the running of the gauge couplings,
and discuss the mixing of the extra $U(1)_{D3}$ of the D3-brane with the Standard Model gauge fields.
Fully specifying the effects of the CFT on the visible sector requires a more detailed
analysis of operator scaling dimensions in cases of non-trivial monodromy,
a task which we defer to future work \cite{HTVW}.

The organization of the rest of this paper is as follows. In section \ref{sec:SEVENBRANE} we
discuss the interplay between the geometry of a local F-theory compactification and the moduli
space of seven-branes. We extend some of the discussion present in the literature, which will
be necessary as preparation for our analysis of D3-brane probe theories. In section \ref{sec:PROBING}
we turn to the worldvolume theory of D3-branes probing an F-theory GUT. In particular, we show how
to translate the seven-brane background fields into the D3-brane probe theory, as well as F-term couplings
between the probe sector and the Standard Model. In section \ref{sec:TRIVIUM} we study
the resulting deformation induced by a probe of trivial seven-brane monodromy, and in section \ref{sec:GAUGE}
we discuss some aspects of how the gauge fields of the Standard Model couple to the probe sector. Section
\ref{sec:CONCLUSIONS} contains our conclusions.

\section{Seven-Brane Gauge Theory and F-theory}\label{sec:SEVENBRANE}

In this section we briefly review the primary ingredients of seven-brane gauge theories
which enter into an F-theory GUT following the discussion in \cite{BHVI}.
We also extend some of the discussion present in the existing literature, in preparation
for our applications to the case of D3-brane probes.

Our starting point is a compactification of F-theory down to four dimensions. Assuming $\mathcal{N} = 1$ supersymmetry in
four dimensions, this data is specified by a choice of an elliptic Calabi-Yau
fourfold which is fibered over a threefold base $B$. The singular fibers of this
geometry determine the locations of seven-branes in the compactification. In addition, a compactification is
specified by a collection of bulk three-form fluxes, as well as fluxes which localize on seven-branes.
Tadpole cancellation conditions also require the presence of D3-branes due to the constraint \cite{Sethi:1996es}:
\begin{equation}
\frac{\chi(CY_{4})}{24} = N_{D3} + \int_{B} H_{RR} \wedge H_{NS}
\end{equation}
where here, $\chi(CY_{4})$ is the Euler character of the Calabi-Yau fourfold,
$N_{D3}$ denotes the number of D3-branes, and the $H$'s denote the three-form
RR and NS fluxes.

The visible sector of the Standard Model is constructed from the intersection of seven-branes in the compactification.
The locations of these seven-branes are dictated by the discriminant locus of the Weierstrass model:
\begin{equation}
y^{2}=x^{3}+fx+g
\end{equation}
where the discriminant of the cubic in $x$ is:%
\begin{equation}
\Delta=4f^{3}+27g^{2}.
\end{equation}
Enhancements in the singularity type of the elliptic fibration dictate the matter content and interactions of the low energy four-dimensional theory.
In the vicinity of the GUT seven-brane, we can introduce a local normal coordinate $z$ such that the
location of the seven-brane then corresponds to the K\"ahler surface $S = (z = 0)$. The gauge group
on the seven-brane is then dictated by the local ADE fibration over $S$. Inside of
the K\"ahler surface $S$, the singularity type can enhance further, giving rise to matter fields
localized on complex curves, and Yukawa interactions localized at the intersection of these curves.
Each of these is accompanied by a further enhancement in the singularity type, yielding the basic
containment relations $G_{S} \subset G_{\Sigma} \subset G_{p}$ for enhancements along a surface $S$,
a complex curve $\Sigma$, and a point $p$. In what follows, we shall often use
the notation $G \equiv G_{p}$.

We can study the matter content and interaction terms in the vicinity of a point $p$
in terms of a partially twisted eight-dimensional gauge theory with gauge group $G_{p}$ \cite{BHVI}.
In this patch, we can model the configuration of intersecting seven-branes in terms of a
background field configuration which Higgses the theory down to a lower singularity type.
The matter content of the seven-brane gauge theory includes a connection $A$ for a
principal $G$-bundle $P$, and an adjoint-valued $(2,0)$ form $\Phi$, which transforms as a
section of $K_{S} \otimes ad(P)$. Internal background field solutions
are specified by a choice $(A,\Phi)$ which satisfy the F-term equations of
motion \cite{VafaWitten,BHVI}:%
\begin{equation}
\overline{\partial}_{A}\Phi=F_{(0,2)}=F_{(2,0)}=0 \label{Fterm}%
\end{equation}
and the D-term equations of motion:%
\begin{equation}
\omega\wedge F_{(1,1)}+\frac{i}{2}\left[  \Phi,\Phi^{\dag}\right]  =0
\label{Dterm}%
\end{equation}
where in the above, $F$ denotes the curvature of the gauge bundle, and
$\omega$ denotes the K\"ahler form of the internal space wrapped by the seven-brane.
The gauge orbit of this field configuration defines a point in the moduli space of the theory.

From the perspective of the four-dimensional field theory, we can organize the seven-brane mode content
as a collection of chiral superfields labelled by the points of the K\"ahler surface $S$. These modes
decompose as a $(0,1)$ component of the gauge field $A$ and the $(2,0)$ form $\Phi$. The breaking
pattern of the gauge theory to the four-dimensional gauge group is specified by the background values $A_{0}$ and $\Phi_{0}$,
and fluctuations around this background determine matter fields:
\begin{align}
A  & =A_{0}+\delta A\\
\Phi & =\Phi_{0}+\delta\Phi.
\end{align}
In principle, the fluctuations $\delta A$ and $\delta \Phi$ can either propagate throughout the
K\"ahler surface $S$, or localize on a curve $\Sigma$ \cite{KatzVafa,BHVI}. In the context of realistic F-theory GUTs,
we typically require that the low energy fluctuations from the bulk are absent so that all matter fields
localize on curves.

An important check on the gauge theory description provided by the seven-brane is how the moduli space of background
field configurations match on to the data of an F-theory compactification. In broad terms, the gauge invariant
data of the seven-brane is characterized in terms of the Casimirs of $\Phi$, and by the gauge field strength of the seven-brane.
One particularly convenient gauge for checking the correspondence between the seven-brane gauge theory, and the
data of an F-theory compactification is in holomorphic gauge. In this gauge, the $(0,1)$ component of the gauge field
is set to zero in a patch, and the equation of motion for $\Phi$ becomes $\overline{\partial} \Phi = 0$.
In terms of local coordinates $z_{1}$ and $z_{2}$ defined on the seven-brane locus,
$\Phi(z_{1},z_{2})$ is holomorphic in the $z_{i}$.  In this gauge, $A_0+\delta A=0$ and {\it all}
the matter fields come from $\delta \Phi$.

Activating a non-zero value for $\Phi(z_{1} , z_{2})$ corresponds to tilting
the seven-branes of the original gauge theory with gauge group $G$.
There is a non-trivial match between the holomorphic
data defined by the Casimirs of $\Phi$, and the ways that we can unfold a
geometric singularity of the F-theory compactification \cite{BHVI}. For example,
the unfolding of an E-type singularity is given as:
\begin{align}
E_{8}  &  :y^{2}=x^{3}+z^{5}+\left(  f_{2}z^{3}+f_{8}z^{2}+f_{14}%
z+f_{20}\right)  x+\left(  g_{12}z^{3}+g_{18}z^{2}+g_{24}z+g_{30}\right) \\
E_{7}  &  :y^{2}=x^{3}+xz^{3}+\left(  f_{8}z+f_{12}\right)  x+\left(
g_{2}z^{2}+g_{6}z^{3}+g_{10}z^{2}+g_{14}z+g_{18}\right) \\
E_{6}  &  :y^{2}=x^{3}+z^{4}+\left(  f_{2}z^{2}+f_{5}z+f_{8}\right)  x+\left(
g_{6}z^{2}+g_{9}z+g_{12}\right)
\end{align}
where the $f_{i}$'s and $g_{i}$'s are degree $i$ polynomials constructed
from the Casimirs of $\Phi$. The position dependence of the Casimirs dictates the loci
of further enhancement in the singularity type, determining where matter fields
localize in the geometry.

Geometrically, the possible ways to unfold a singularity are dictated by elements of the Cartan subalgebra of $G$ modulo the
group action of the Weyl group $W(G)$ on the fundamental weights of $G$ \cite{KatzMorrison}. Often, the breaking pattern is of the form
$G \supset G_{S} \times G_{\bot}$, specifying unfolding to a subgroup $G_S$. In this case,
$\Phi$ takes values only in $G_{\bot}$. The case of maximal interest for us in this paper is given by
unfolding $G = E_8$ via the breaking pattern $E_{8} \supset SU(5)_{GUT} \times SU(5)_{\bot}$,
where $\Phi$ takes values in $SU(5)_{\bot}$. This induces a geometric unfolding to a bulk $SU(5)$ seven-brane
gauge theory, with matter curves specified by the profile of $\Phi$.

Given a local description of a Calabi-Yau fourfold, it is also natural to ask how this data is encoded in the seven-brane gauge theory.
This is not enough information to reconstruct a unique answer. The reason is that in the seven-brane moduli space,
the Casimirs of $\Phi$ and the flux data are what specifies a field configuration. Assuming that $\Phi$ takes values in the Cartan subalgebra of $G$,
we can pass back and forth between the Casimirs of $\Phi$, and $\Phi$ itself.  However, equation (\ref{Dterm})
allows for the more general possibility of $\Phi$ not being in the Cartan.  This means that there
may be loci where we cannot diagonalize $\Phi$.  Diagonalizing
$\Phi$ away from such loci, it may
happen that the eigenvalues of $\Phi$ exhibit branch cuts in the variables $z_1$ and $z_2$. This is the phenomenon of
seven-brane monodromy \cite{Hayashi:2009ge} (see also \cite{BHSV,EPOINT,Marsano:2009gv}).

We can also characterize this branch cut structure in terms of Higgs
bundle data \cite{HitchinSelf,SimpsonGeneralize}, which has been discussed in
the context of F-theory compactifications for example in \cite{DWIII}. At a generic point of the K\"ahler surface,
we introduce a field $\Phi$ taking values in the Cartan subalgebra of $G$. As
we move from point to point, it may happen that the basis in which $\Phi$ evaluated at this point
appears diagonal may be different. Defining a compact
holomorphic curve $C$, passing $\Phi$ around this curve and back to the same point $p$ may transform the value
of $\Phi$ at the point $p$ as:
\begin{equation}\label{Berry}
\Phi(p) \rightarrow g^{-1}(p) \Phi(p) g(p)
\end{equation}
where the adjoint action by $g(p)$ in the complexified gauge group amounts
to permuting the eigenvalues of $\Phi(p)$. This provides an
equivalent characterization of seven-brane monodromy.

Strictly speaking, $\Phi$ defined in this way is single-valued
only after deleting the branch cut locus for its eigenvalues. Indeed,
as a particle traverses a branch cut, it experiences a non-abelian Berry phase, transforming
as in equation (\ref{Berry}). To establish the presence
of the flux which induces this Berry phase, we now consider
the field configuration defined by $A$ and $\Phi$ on the patch with the branch
loci deleted. Since we are away from the branch locus, there exists a smooth
gauge transformation on this patch by an element $g(z_{1} , z_{2})$ of
the complexified gauge group:
\begin{align}\label{gaugetransf}
A  & \rightarrow g^{-1}A g -g^{-1}dg\\
\Phi & \rightarrow g^{-1}\Phi g
\end{align}
such that in the gauge transformed presentation, the position dependence of
$\Phi(z_{1},z_{2})$ is smooth in the $z_{i}$, and in particular exhibits no
branch cuts. Though $g(z_{1},z_{2})$ will be smooth at a generic point in this
patch, along the branch cut locus, this gauge transformation will be singular,
and signals the presence of a Dirac string for the gauge field. Indeed, after
performing this gauge transformation, $\Phi$ will no longer be diagonal, and
$\left[  \Phi,\Phi^{\dag}\right]  \neq0$. To satisfy the D-term equations of
motion, this requires non-zero $F_{(1,1)}$, establishing the
presence of a background gauge field flux. To summarize, in the gauge where no
branch cuts are present in $\Phi$, a gauge field flux will be spread out over
the entire patch. In the gauge with branch cuts for $\Phi$, there will instead
be a Dirac string of flux localized along the (deleted) branch locus.

As an example of seven-brane monodromy, consider a background field
configuration with $\Phi=\phi$ $dz_{1}\wedge dz_{2}$ with:%
\begin{equation}
\phi=\left[
\begin{array}
[c]{cc}%
0 & 1\\
z_{1} & 0
\end{array}
\right]  .\label{phimono}%
\end{equation}
The eigenvalues of $\phi$ are $\pm\sqrt{z_{1}}$, indicating the presence of a
branch cut along $z_{1}=0$, and $\mathbb{Z}_{2}$ seven-brane monodromy.
Let us note that in equation (\ref{phimono}), the quadratic Casimir
$Tr(\phi^{2})=2z_{1}$ is generically non-zero, though it vanishes at $z_{1} = 0$.
Note, however, that this does not mean that $\phi$ vanishes at this point.
Rather, it has become a nilpotent matrix. This will be important later when we discuss such deformations
from the perspective of the probe theory.\footnote{As a brief aside, let us note that though the matrices
$\phi_{1}=\left[
\begin{array}
[c]{cc}%
z_{1} & 0\\
0 & -z_{1}
\end{array}
\right]$
and
$\phi_{2}=\left[
\begin{array}
[c]{cc}%
0 & 1\\
z^{2}_{1} & 0
\end{array}
\right]$ have the same eigenvalues, they correspond to different geometries, in the sense that they define
different moduli spaces of possible deformations. Indeed, under a small perturbation in the non-zero matrix
entries, we see that $\phi_{1}$ retains its general form, whereas the eigenvalues of $\phi_{2}$ develop non-trivial branch cuts. Thus,
although these matrices have the same Casimirs, the latter case is not generic, and we shall not consider this case further in what
follows.}

In most applications we consider in this paper, we shall work in terms of the smooth $\Phi$ obtained by
performing a complexified gauge transformation on the $\Phi$ configuration which exhibits branch cuts. One
reason for doing this is that such field configurations are manifestly holomorphic, and all quantities
of the seven-brane gauge theory are then non-singular. The other reason for this choice is that in the context of D3-brane probe
theories, the $z_i$ will be promoted to the vevs of fields $Z_i$. Taking a root of a field is a rather ill-defined
notion in field theory, and signals that the appropriate gauge choice is one in which $\Phi$ is analytic in
the $z_i$.

\section{Probing an F-theory GUT}\label{sec:PROBING}

In the previous section we discussed the interplay between the moduli space of
the eight-dimensional gauge theory, and the unfolding of an F-theory singularity.
We now turn to the worldvolume theory of a D3-brane probing this
configuration of intersecting seven-branes.

This section is organized as follows. We first review $\mathcal{N} = 2$ D3-brane
probes of seven-branes. These probe theories correspond to a strongly
interacting theory of the type studied by Minahan and Nemeshansky plus a decoupled free
hypermultiplet. The more general case of $\mathcal{N} = 1$ probe theories correspond to deformations by operators
which couple these two theories together.

\subsection{Review of $\mathcal{N} = 2$ Probes}

To frame the discussion to follow, we now review the worldvolume theory of D3-brane probes which preserve $\mathcal{N} = 2$ supersymmetry.
Let us first consider the theory of the D3-brane away from all seven-branes
of the compactification. In a sufficiently small patch of the threefold base, the geometry
probed by the D3-brane is $%
%TCIMACRO{\U{2102} }%
%BeginExpansion
\mathbb{C}
%EndExpansion
^{3}$, and the position of the D3-brane is parameterized by the chiral
superfields $Z_{1}$, $Z_{2}$ and $Z$.
The holomorphic $\tau_{YM}$ of the D3-brane is
specified by the IIB $\tau$, which is in turn determined
implicitly by the elliptic curve:
\begin{equation}
y^2 = x^3 + fx + g
\end{equation}
via the corresponding $j$-function:
\begin{equation}
j(\tau)=\frac{4(24f)^{3}}{4f^{3} + 27 g^{2}}.
\end{equation}

We now turn to the theory of the D3-brane in the neighborhood of a stack
of parallel seven-branes with gauge group $G$. We parameterize the local
geometry in terms of a coordinate $z$ normal to the seven-brane such that the
seven-brane is located at $z=0$. Here, $z_{1}$ and
$z_{2}$ are coordinates parallel to the seven-brane. In the limit where
the seven-brane worldvolume is non-compact, this gauge group corresponds to a
flavor symmetry of the D3-brane theory. In realistic applications, we can view
$G$ as a flavor group which has been weakly gauged by compactifying the
eight-dimensional gauge theory of the seven-brane.

The moduli space of the D3-brane probe has a Coulomb branch and a Higgs
branch. The Coulomb branch describes motion away from the seven-brane. This is
parameterized by the vev $z = \langle Z \rangle$. At the origin of the Coulomb branch
the D3-brane sits on top of the seven-brane. At this
point, the original D3-brane can dissolve into flux inside the seven-brane
which corresponds to the Higgs branch of the D3-brane worldvolume theory. When
the seven-brane gauge group $G$ is of $SU$, $SO$ or $USp$ type, the
ADHM\ construction establishes that the Higgs branch corresponds to the moduli
space of instantons of the internal four-dimensional gauge theory of the
seven-brane. For exceptional gauge groups, there is no known
ADHM construction of instantons. Nevertheless, the physical picture of
D3-branes dissolving into flux still provides a qualitative way to view the
Higgs branch.

At $z=0$, additional light states enter the worldvolume theory of the
D3-brane. In F-theory, these correspond to $(p,q)$ strings and their
junctions which stretch between the seven-brane and the D3-brane. For
perturbatively realized configurations such as seven-branes with gauge group
$SU(n)$, these are $3-7$ bifundamental strings. For E-type gauge groups, the
exceptional seven-brane is a bound state of seven-branes of different $(p,q)$
types, and so we can expect both electric and magnetic states of different spins
to enter the low energy theory. These states cause the gauge coupling of the D3-brane to run in
different directions. For example, the probe of the $E_{8}$ singularity:%
\begin{equation}
y^{2}=x^{3}+z^{5}%
\end{equation}
has vanishing $j$-function, indicating that the coupling does not run as a
function of $z$, and remains fixed at $\tau=\exp(2\pi i/3)$.

The operators of the probe theory transform in representations of the seven-brane
group $G$, which corresponds to a flavor symmetry of the D3-brane worldvolume theory. In
the weakly coupled setting, these operators can be viewed as composite
operators constructed from more basic fields. For example, in the D3-brane
probe theory of a seven-brane with gauge group $SU(n)$, the $3-7$ strings form
vector-like pairs of quarks and anti-quarks $Q_{i}\oplus\widetilde{Q}_{i}$.
These quarks are bifundamentals, charged under both the seven-brane group $G$
and the D3-brane gauge group $U(1)$. The mesonic branch of the moduli space is
parameterized in terms of the $U(1)$ gauge invariant combination:%
\begin{equation}
M_{i\overline{j}}=Q_{i}\widetilde{Q}_{\overline{j}}%
\end{equation}
which transforms in the adjoint representation of $SU(n)$. The $Q$'s and
$\widetilde{Q}$'s parameterize the instanton moduli space via the
ADHM construction \cite{ADHM}, and its string theoretic
description \cite{WittenADHM,DougInstI,DougInstII}.

In more general settings, we shall be interested in the analogue of the
mesonic operators for D3-brane probe theories of E-type singularities. Here,
the presence of both electric and magnetic states means that we should not
expect a weakly coupled Lagrangian formulation which contains both the quarks
and their magnetic duals. We can, however, still work in terms of operators
which are the analogues of the mesonic operators, which will transform in the adjoint
representation of the flavor symmetry group $G$.

E-type singularities are of particular importance for F-theory GUTs, and will be our main
focus in this paper. The Seiberg-Witten curve for the $\mathcal{N}=2$ rank one superconformal field
theories with exceptional flavor symmetry $E_{n}$ are given by the equations
for the corresponding E-type singularity \cite{MNI,MNII}:\footnote{Here the rank
of a SCFT denotes the dimension of the Coulomb branch.}%
\begin{align}
E_{8}  &  :y^{2}=x^{3}+z^{5}\\
E_{7}  &  :y^{2}=x^{3}+xz^{3}\\
E_{6}  &  :y^{2}=x^{3}+z^{4}\text{.}%
\end{align}
In the context of F-theory, these theories are realized by D3-brane probes of
the corresponding E-type seven-brane sitting at $z=0$. More precisely, the
D3-brane worldvolume theory consists of the $E_{n}$ SCFT, and a free
$\mathcal{N}=2$ hypermultiplet which describes motion parallel to the
seven-brane. Thus, the D3-brane probe is the direct sum of two decoupled CFTs,
$CFT(E_{n})\oplus CFT_{free}$. The spin zero chiral primaries of $CFT(E_{n})$ are $Z$,
and an operator $O$ transforming in the adjoint representation
of $E_{n}$. The chiral primary of $CFT_{free}$ is given by the
single hypermultiplet $Z_{1}\oplus Z_{2}$. The scaling dimensions of these operators
for the various D3-brane probe theories are:%
\begin{equation}%
\begin{tabular}
[c]{|c|c|c|c|c|}\hline
& $[Z_{1}]$ & $[Z_{2}]$ & $[Z]$ & $[O]$\\\hline
$E_{6}$ & $1$ & $1$ & $3$ & $2$\\\hline
$E_{7}$ & $1$ & $1$ & $4$ & $2$\\\hline
$E_{8}$ & $1$ & $1$ & $6$ & $2$\\\hline
\end{tabular}
\text{ \ \ }.
\end{equation}
The counting of electric states in the E-type theories has been studied in
string theory in \cite{Klemm:1996hh}, and an index for the $E_6$
probe theory has been determined in \cite{Gadde:2010te}. Though these theories do
not possess a weakly coupled Lagrangian formulation, there do exist
gauge theory duals \cite{Argyres:2007cn,Argyres:2007tq}.

Mass deformations of the original $\mathcal{N} = 2$
are parameterized in terms of the F-term deformations
\begin{equation}
\delta L = \int d^{2} \theta \,\, Tr_{G}(m \cdot O) + h.c.
\end{equation}
where in the above, both $m$ and $O$ transform in the adjoint representation of $G$. In order
for this deformation to preserve $\mathcal{N} = 2$ supersymmetry, we must require that
$[m , m^{\dag}] = 0$, much as in the weakly coupled examples studied in
\cite{Argyres:1996eh}. This follows upon weakly gauging the flavor symmetry group $G$, since the
D-term constraint then requires that this commutator vanishes. Let us note that
in the context of generalizations common to F-theory, this commutator does not need
to vanish and the D-term equation is instead
satisfied through the presence of a background flux term, which breaks
$\mathcal{N} = 2$ to $\mathcal{N} = 1$ supersymmetry.

At the level of the $\mathcal{N} = 2$ curve, these deformations are parameterized in terms of
the Casimirs of $m$:
\begin{align}
E_{8}  &  :y^{2}=x^{3}+z^{5}+\left(  f_{2}z^{3}+f_{8}z^{2}+f_{14}%
z+f_{20}\right)  x+\left(  g_{12}z^{3}+g_{18}z^{2}+g_{24}z+g_{30}\right) \\
E_{7}  &  :y^{2}=x^{3}+xz^{3}+\left(  f_{8}z+f_{12}\right)  x+\left(
g_{2}z^{2}+g_{6}z^{3}+g_{10}z^{2}+g_{14}z+g_{18}\right) \\
E_{6}  &  :y^{2}=x^{3}+z^{4}+\left(  f_{2}z^{2}+f_{5}z+f_{8}\right)  x+\left(
g_{6}z^{2}+g_{9}z+g_{12}\right)
\end{align}
where the $f_{i}$ and $g_{i}$ are degree $i$ polynomials built from holomorphic expressions in $m$ which
are neutral under the flavor symmetry. From the perspective of F-theory, this is to be expected. Indeed, we have
already seen that the Casimirs of the adjoint-valued Higgs field $\Phi$ also parameterize deformations of the
singularity type. Introducing the decomposition:
\begin{equation}
\Phi = \phi \,\, dz_1 \wedge dz_2,
\end{equation}
we see that the mass deformations we have been considering correspond to a specific, locally constant
$\phi$ taking values in the Cartan subalgebra. In this simple case, we see that the parameters of the seven-brane
gauge theory enter the probe D3-brane theory as the F-term deformation:
\begin{equation}\label{Odef}
\delta L = \int d^{2} \theta \,\, Tr_{G}(\phi \cdot O) + h.c..
\end{equation}

In the case of perturbative IIB strings, there is an
analogous set of deformations for D3-brane probes of an $SU(n)$ seven-brane.
Indeed, in that context, we can write a local $A_{n-1}$ singularity as:%
\begin{equation}
y^{2}=x^{2}+z^{n}%
\end{equation}
and deformations of the singularity are parameterized by Casimirs of $\phi$
via the deformations:%
\begin{equation}
y^{2}=x^{2}+z^{n}+\left(  \underset{i=2}{\overset{n}{%
%TCIMACRO{\dsum }%
%BeginExpansion
{\displaystyle\sum}
%EndExpansion
}}\sigma_{i}z^{n-i}\right)
\end{equation}
where the $\sigma_i$ are the elementary symmetric
polynomials in the eigenvalues of $\phi$. Letting
$Q_{i}\oplus\widetilde{Q}_{i}$ for $i=1,...,n$ denote the $3-7$
strings corresponding to quarks and anti-quarks of the D3-brane theory, the
analogue of equation (\ref{Odef}) is given by:%
\begin{equation}
\delta L = \int d^{2}\theta\text{ }Tr_{SU(n)}\left(  \phi\cdot Q\widetilde {Q}\right) + h.c.
\end{equation}
where $Q\widetilde{Q}$ is the analogue of $O$, and transforms in the adjoint
representation of $SU(n)$. In other words, activating a vev for $\phi$ induces a
mass for some of the quarks. This is physically reasonable since moving a
seven-brane off of the D3-brane induces a mass for the corresponding pair of
$3-7$ strings.

\subsection{$\mathcal{N}=1$ Probes}\label{ssec:N1}

We now turn to D3-brane probes of intersecting seven-brane
configurations. This corresponds to taking the original parallel stack of seven-branes with gauge group $G$,
and tilting these seven-branes off of each other by turning on a non-zero position dependent vev:
\begin{equation}
\Phi_0(z_{1} , z_{2}) = \phi_0(z_1 , z_2) \,\, dz_1 \wedge dz_2.
\end{equation}
In this case, there will still be loci where the eight-dimensional gauge
symmetry is restored, dictated by the position dependence of
$\Phi_0(z_1 , z_2)$.

From the perspective of the probe D3-brane, the main difference from the case of $\mathcal{N} = 2$
mass deformations considered previously is that now, these masses have some dependence on the previously decoupled CFT
given by the hypermultiplet $Z_1 \oplus Z_2$.  At the level of F-terms, we therefore see that this
translates to the F-term deformation:
\begin{equation}
\delta L = \int d^{2} \theta \,\, Tr_{G} (\phi_{0}(Z_{1} , Z_{2}) \cdot O) + h.c..
\end{equation}
Let us note that the most general possible deformation by $\phi_0$ will not
take values in the Cartan subalgebra. In particular, we see that although it is common in studies of
seven-brane monodromy to work in terms of a diagonalized $\phi_0$ with some choice of branch cuts,
this is inappropriate for the probe D3-brane theory. The absence of a diagonalizable $\phi_0$ also means that
in the D-term equations of motion, there will be a non-zero gauge field flux switched on along the directions of the seven-brane.
It is therefore important to include the effects of this background flux on the probe D3-brane theory.

But as explained for example in \cite{FGUTSNC}, gauge field fluxes do not deform
B-brane superpotential terms. This is because in a sufficiently small patch, we can always
work in a gauge in which the $(0,1)$ component of the gauge field has been set to zero.
In the four-dimensional effective field theory, the $(0,1)$ contribution to the gauge field corresponds to a
collection of chiral superfields \cite{BHVI}. We therefore conclude that such terms do not deform the superpotential. On the other
hand, we do generically expect some contribution from the $(1,1)$ component of the field strength to the
D-terms of the probe theory. Letting $J(F_{(1,1)})$ denote the corresponding D-term source, with possible dependence
on the operators of the original $\mathcal{N} = 2$ theory, the deformation of the theory is:
\begin{equation}
\delta L = \int d^{4} \theta \,\, J(F_{(1,1)}) + \int d^{2} \theta \,\, Tr_{G}(\phi_{0}(Z_1 , Z_2) \cdot O) + h.c..
\end{equation}
From the perspective of the D3-brane probe, all of these contributions are finite deformations of the theory. As all finite
D-term deformations correspond to irrelevant operators, we see that the only source of relevant and marginal deformations
descend from the F-terms, and in particular, from $\phi_{0}(Z_1 , Z_2)$ and its coupling to the operator $O$. The relevant and marginal
deformation of the theory is given by the superpotential term:
\begin{equation}
\delta W = Z_{1} \times Tr_{G}\left(  \lambda_{1}\cdot O\right)  +
Z_{2} \times Tr_{G}\left(  \lambda_{2}\cdot O\right)  + Tr_{G}\left(  m \cdot O\right)
\end{equation}
where the $\lambda$'s correspond to constant matrices transforming in the
adjoint representation of the flavor symmetry group $G$. In other
words, starting from the $\mathcal{N}=2$ system $CFT(E_{n})\oplus CFT_{free}$,
we have added a deformation which couples these two CFTs. Here, we have
included possible contributions from both UV marginal operators such as
$Z_{i}\times Tr (\lambda_i \cdot O)$, as well as relevant deformations such as $Tr( m \cdot O)$. The condition
that we retain a point of $G$ enhancement at the origin of the geometry $z_1 = z_2 = 0$
requires that all of the Casimirs of $m$ are trivial.

The case of trivial seven-brane monodromy where we enhance back to $G$
at the Yukawa point corresponds to the case where the $\lambda$'s take
values in the Cartan subalgebra of $G$, and $m = 0$. The case of non-trivial
seven-brane monodromy is characterized by matrices which are not
simultaneously diagonalizable. For example, the field configuration of equation \eqref{phimono}
corresponds to a combination of a relevant and a marginal deformation of the original $\mathcal{N} = 2$
theory.

\subsection{Coupling to the Visible Sector}

In our discussion so far, we have treated the fields of the seven-brane as background parameters. In
realistic applications, we must compactify this system, and the matter fields become dynamical. Note, however, that
we have already deduced the general coupling of $\Phi = \phi \,\, dz_1 \wedge dz_2$ to the D3-brane probe:
\begin{equation}
\delta L = \int d^{2} \theta \,\, Tr_{G}(\phi(Z_{1} , Z_{2}) \cdot O) + h.c..
\end{equation}
Using the general expansion of $\Phi$ into a background contribution and its fluctuations:
\begin{equation}
\Phi = \Phi_{0} + \delta \Phi
\end{equation}
we conclude that the matter fields of the visible sector couple to the Standard Model via:
\begin{equation}
\delta L = \int d^{2} \theta \,\, Tr_{G}(\phi_{0}(Z_{1} , Z_{2}) + \delta \phi(Z_{1} , Z_{2}) \cdot O) + h.c..
\end{equation}
In the above expression, $\delta \phi$ is shorthand for the matter field fluctuations which can either
propagate in the bulk of $S$, or localize on a matter curve.

Since it is the case of primary interest
for realistic F-theory GUTs, let us consider the special case where $\delta \phi$ describes a six-dimensional field.
Specifying the explicit form of the couplings between matter fields and the D3-brane probe would require
a more complete study of the profile of matter field wave functions with non-trivial seven-brane monodromy,
a task which is beyond the scope of the present paper. In what follows, we therefore restrict attention to the
non-monodromic case, though we expect similar formulae to hold more generally.

Four-dimensional fields localized on matter curves are
given by holomorphic sections of a line bundle defined
over the matter curve. Introducing a local coordinate $z_{\parallel}$
along the curve, and a coordinate normal to the curve $z_{\bot}$, the profile of the four-dimensional
wave functions are, in holomorphic gauge, given by a power series in $z_{\parallel}$, which
we organize according to their order of vanishing near the Yukawa point:
\begin{equation}
\delta \phi_{R}=\underset{g}{%
%TCIMACRO{\dsum }%
%BeginExpansion
{\displaystyle\sum}
%EndExpansion
}h_{g}(z_{\parallel}) \psi_{R}^{(g)} + (\text{massive modes}).
\end{equation}
Here, the four-dimensional field transforms in a representation $R$ with respect to the
gauge group left unbroken by the unfolding of the singularity $G$. In most applications where we decompose
$E_8 \supset SU(5)_{GUT} \times SU(5)_{\bot}$, we can further decompose $R$ into a representation of $SU(5)_{GUT}$,
and a representation of the subgroup of $SU(5)_{\bot}$ left unbroken by the geometric unfolding.
The field $\psi_{R}^{(g)}$ denotes a massless generation of
the Standard Model. The expression $h_{g}(z_{\parallel})$ corresponds to a power series in
$z_{1}$ and $z_{2}$ such that for the heaviest generation, $h_{3}$ does not
vanish at $z_{1}=z_{2}=0$, and for the lighter generations, there is a higher
order of vanishing for the $h_{i}$.

Promoting the coordinate dependence in $h_{g}$ to a field dependent wave function profile,
we therefore deduce that the visible sector couples to the D3-brane probe through the couplings:
\begin{equation} \label{37coupling}
W_{3-7} = \underset{g}{%
%TCIMACRO{\dsum }%
%BeginExpansion
{\displaystyle\sum}
%EndExpansion
} h_{g}(Z_{\parallel})\psi_{R}^{(g)}\cdot O_{R^{\ast}}
\end{equation}
where here, we have decomposed the operator $O$ which transforms in the adjoint of $G$
into irreducible representations of the group left unbroken by the unfolding, such that $O_{R^{\ast}}$
transforms in the representation dual to $R$. Specifying all details of the matter field couplings requires us
to determine more details of how seven-brane monodromy acts on the matter field wave functions.

It is also natural to expect that the matter fields of the Standard Model will
couple to the $3-7$ strings through additional higher dimension operators.
Indeed, integrating out massive modes of the compactification, we can expect
on general grounds superpotential couplings of the form:%
\begin{equation}
\delta W = \phi^{m}\cdot O^{n} .
\end{equation}
It would be interesting to systematically estimate the form of all such couplings.

\subsection{Flux and D3-Branes}

As noted in \cite{FGUTSNC} in order to minimally realize
flavor hierarchies, we need to have a suitable flux turned on.
In the presence of this flux the F-term equations of
motion for $n$ coincident D3-branes at a generic point of the
seven-brane are:
\begin{equation}\label{NCmotion}
\left[  Z_{i}, Z_{j}\right]  =\theta^{ij}(\overrightarrow{Z}),
\end{equation}
where $\theta^{ij}$ is set by the flux.
As studied for example in \cite{Martucci:2006ij} and \cite{FGUTSNC}, 
this equation of motion is obtained from a flux-induced superpotential term:%
\begin{equation}\label{WNC}
W= \varepsilon_{ijk} Tr_{U(n)} \left( Z_{i} Z_{j} Z_{k} + \int ^{Z_{k}}\theta^{ij}(\overrightarrow{Z}) dZ_{k} \right).
\end{equation}
Favorable flavor hierarchies require $\theta^{ij}$ to vanish at the Yukawa point \cite{FGUTSNC}, so that D3-branes
are naturally attracted to the Yukawa point. In the case of a single D3-brane, the usual $\mathcal{N} = 4$ superpotential term
vanishes, and we are left with only the second term of equation (\ref{WNC}).
In that case, to leading order this gives a term of the form:
\begin{equation}
W=(\alpha Z_1 +\beta Z_2)\cdot Z,
\end{equation}
where as in the previous sections, $Z_1,Z_2$ denote directions
parallel to the seven-brane and $Z_3=Z$ is the direction normal to the seven-brane.
Note that in the original $\mathcal{N}=2$ CFT, these terms are irrelevant,
as the dimension of $Z$ is at least three. We will see later in the paper that this
continues to be the case when we have trivial monodromy.

\section{Probing an E-point}\label{sec:TRIVIUM}

In this section we study the resulting $\mathcal{N} = 1$ theory obtained by probing
an E-type Yukawa point. In this case, the original $\mathcal{N} = 2$ theory has flavor symmetry
$G = E_{n}$, and the $\mathcal{N} = 1$ theory is obtained by a deformation of this theory.
Though a full analysis of $\mathcal{N} = 1$ deformations of the original $\mathcal{N} = 2$ theories
is beyond the scope of the present paper, in the special case of trivial seven-brane monodromy
we can determine the resulting low energy dynamics. An analysis of the IR theories
resulting from non-trivial seven-brane monodromy will be given elsewhere \cite{HTVW}.

We now show that in a probe of trivial seven-brane monodromy,
the corresponding deformation of the $\mathcal{N} = 2$ theory
is marginal irrelevant, and therefore induces a flow back to the \textit{original} theory!
To establish this, first recall that trivial seven-brane monodromy means
$\phi_{0}(z_1 , z_2)$ takes values in the Cartan subalgebra of the
flavor symmetry group $G$. In such cases, the deformation does not
include a D-term contribution, and is fully characterized by the superpotential deformation:
\begin{equation}
\delta W = Z_{1} \times Tr_{G}\left(  \lambda_{1}\cdot O\right)  +
Z_{2} \times Tr_{G}\left(  \lambda_{2}\cdot O\right)
\end{equation}
in which $\lambda_1$ and $\lambda_2$ both lie in the Cartan subalgebra of $G$.

To study the effects of this deformation we can apply the general result of
\cite{Green:2010da}, which provides a group-theoretic characterization of exactly
marginal deformations of a conformal theory. The basic result in \cite{Green:2010da} is
that we can classify the space of marginal deformations by weakly gauging all
of the flavor symmetries $G_{\rm{Flavor}}$ of the system, and their action on the space of
couplings $\left\{  \lambda\right\}  $. Performing the symplectic quotient:%
\begin{equation}
\mathcal{M}_{\rm{couplings}} = \left\{  \lambda\right\}  \,\, // \,\, G_{\rm{Flavor}}
\end{equation}
then yields the space of exactly marginal couplings. In the present context,
we can re-write the original deformation as:%
\begin{equation}
\delta L=\int d^{2}\theta \,\, Tr_{G}\left(  \left[
\begin{array}
[c]{cc}%
\lambda_{1} & \lambda_{2}
\end{array}
\right]  \left[
\begin{array}
[c]{c}%
Z_{1}\\
Z_{2}%
\end{array}
\right]  \cdot O\right) + h.c..
\end{equation}
Thus, we see that the couplings $\lambda_1$ and $\lambda_2$ transform as a doublet
under $U(2)\subset G_{\rm{Flavor}}$ rotations. Note in particular that both components have the same
charge under the $U(1)$ in the center of $U(2)$. This means that the \textquotedblleft
D-term constraint\textquotedblright\ of the symplectic quotient identifies all
couplings with the case where there is zero deformation. In other words, we
learn that the space of exactly marginal couplings is trivial. Since UV marginal
operators are either exactly marginal or marginal irrelevant \cite{Green:2010da},
it follows that the original deformation induces a flow back to the original
$\mathcal{N}=2$ theory. In particular, all of the operators have the same scaling dimension
as they had in the UV.

In the case of non-trivial seven-brane monodromy, it is more difficult to track the infrared behavior of the theory,
in part because it will be a combination of relevant and marginal deformations which are not diagonalizable. In general,
it is a difficult task to prove that the infrared dynamics induces a flow to a CFT. The main assumption
we shall implicitly make is that the $\mathcal{N} = 1$ deformations we consider here induce
flows to non-trivial interacting superconformal field theories. Our evidence for this
is circumstantial, but also self-consistent. First, we have already observed that with trivial seven-brane monodromy,
probes of E-points induce a flow back to the original $\mathcal{N} = 2$ theory. Second, the presence of an
exceptional singularity indicates that light electric and magnetic states will always be present in the corresponding
probe theory. This is a non-trivial indication that an interacting theory of some sort is present at this point.
Third, we can compute the value of the dilaton as we approach the E-point. Though this depends on the path of approach,
there always exists a path along which the dilaton is constant. This again provides a hint of interesting behavior at the origin of
the Coulomb branch.

In this section we have ignored the effect of couplings of the brane probe
to the degrees of freedom on the seven-brane, which is valid in the limit
$\alpha_{GUT}\rightarrow 0$.\footnote{Recall that $\alpha_{GUT}$ controls
the inverse volume of the K\"ahler surface wrapped by the seven-brane.} Taking into account these couplings
would weakly gauge the corresponding flavor symmetries descending from the seven-brane, as well as
introduce additional couplings to the matter sector noted before.

\section{Coupling to Gauge Fields}\label{sec:GAUGE}

Up to this point, our discussion has focussed on some of the basic
features of how a D3-brane probe of an exceptional point would
couple to the matter fields of the Standard Model. In this section we discuss
how this theory couples to the Standard Model gauge fields.

As a very basic point, let us note that the conformal symmetry of the D3-brane
probe must already be broken at energy scales of a few hundred GeV. The reason is that in all cases, the probe
D3-brane contains $3-7$ strings charged under both the Standard Model gauge group and the D3-brane
probe theory, as reflected for example in the operators $O$. In order to have avoided detection thus far,
it is therefore necessary to assume that all such charged states are sufficiently heavy. Placing the
dynamics of the probe theory at a few hundred GeV or higher is also natural
in the sense that whatever dynamics is responsible for breaking supersymmetry will also induce some potential for the
degrees of freedom of the probe theory.

The fact that $M_{\cancel{CFT}}$ is bounded below by a few hundred GeV is quite
different from the unparticle scenario considered
in \cite{Georgi:2007ek}, with its approximately conformal
sector at the weak scale. Rather, at the energy scale $M_{\cancel{CFT}}$, there will be
additional particle states which enter the low energy theory.
Proceeding up to sufficiently high energy scales, we can expect additional massive
states of different spins to contribute to the theory. At sufficiently high energy scales, the
theory is better described as a conformal field theory, and scale invariance in this sector is approximately restored.
The possible applications of the D3-brane probe theory depend somewhat on the energy scale $M_{\cancel{CFT}}$. While
a lower scale of conformal symmetry breaking is phenomenologically quite interesting, one can in principle envision
$M_{\cancel{CFT}}$ being much higher, and this may be of interest for other model building applications.

In the remainder of this section we study some of the ways that the D3-brane probe theory
interacts with the gauge fields of the Standard Model. While the most spectacular consequences
would come from the conversion of gauge fields into TeV scale $3-7$ strings, it is also more difficult
to extract quantitative information about the phenomenology of this scenario, a task which we defer
to future work \cite{WorkProgHV}.

One question we can address, however, is the effect of this sector on the unification of gauge couplings. At sufficiently
high scales where the theory is approximately conformal, a tower of charged particles
will enter the spectrum, which might appear to pose problems for perturbative
gauge coupling unification. Even though there are a large number of states
contributing to the running of the gauge coupling, we argue that the effect on the running is far milder,
and retains perturbative gauge coupling unification.

The probe theory also contains a $U(1)_{D3}$ gauge sector of its own, which can interact with the Standard Model via
kinetic mixing with $U(1)_{Y}$. A novel feature of this type of theory is that
generically, there can be kinetic mixing involving both the
electric and magnetic dual field strengths of this extra $U(1)_{D3}$.

\subsection{Current Correlators}

In this subsection we compute the effects of the D3-brane probe theory on the running of the
Standard Model gauge couplings. More precisely, we compute the effects from the probe theory
in the regime where the D3-brane probe is approximately conformal. In this regime, a number
of additional states charged under $SU(3)_C \times SU(2)_L \times U(1)_Y$ enter as threshold
corrections. Moreover, these states interact strongly with the conformal sector of the D3-brane probe.
It is therefore important to check that the presence of these states do not spoil gauge coupling unification,
and moreover, do not induce a Landau pole at low scales. Proceeding from low energies near $M_{\cancel{CFT}}$ to the scale
where conformal symmetry is approximately restored, our expectation is that there is some complicated interpolation
which takes account of these various thresholds.

Even though we are dealing with a strongly coupled CFT, note that
all of the electric and magnetic states descend from $(p,q)$ strings and their junctions
which fill out complete GUT multiplets. In particular, this means
that the contribution from the probe sector will preserve gauge coupling unification,
and its scale. For this reason, it is enough to phrase our discussion in terms of the effects of
the probe on the running of the $SU(5)$ gauge coupling.

We now study the running of the gauge coupling constant due to the D3-brane probe theory.
The computation we present exploits the overall holomorphy present in the gauge coupling constants. In a weakly
coupled gauge theory such as the Standard Model, we can use this result to
extract the one loop running of the physical gauge coupling constant.

Following \cite{GanorKrogh}, in more formal terms, computing the effects on the
gauge coupling from the probe D3-brane amounts to computing the current
correlator for the flavor symmetry of the $\mathcal{N}=1$ SCFT theory:%
\begin{equation}
\left\langle J_{\mu}^{A}(q)J_{\nu}^{B}(-q)\right\rangle =\rho \cdot \delta
^{AB}\left(  q^{2}\eta_{\mu\nu}-q_{\mu}q_{\nu}\right)  \times\left\{
\begin{array}
[c]{c}%
\log\frac{\Lambda}{M}\text{ for }\left\vert q\right\vert \ll M\\
\log\frac{\Lambda}{\left\vert q\right\vert }\text{ for }\left\vert
q\right\vert \gg M
\end{array}
\right\}  . \label{currentcorrelator}%
\end{equation}
where $A$ and $B$ are indices in the adjoint representation of the flavor
group, $\Lambda$ is a cutoff of the field theory, and $M$ is a characteristic
mass scale. Weakly gauging the flavor symmetry, the current correlator
determines the running of this gauge coupling constant as a function of energy
scale:%
\begin{equation}
\frac{1}{g^{2}\left(  \mu\right)  }=\frac{1}{g^{2}\left(  \Lambda\right)
}-\rho\log\left(  \frac{\Lambda}{\mu}\right)  \text{.}%
\end{equation}

Our strategy for extracting the current correlators will be to perturb the
probe theory to an $\mathcal{N}=1$ system in which the gauged flavor symmetry
admits a weakly coupled description. Computing the running of the couplings in
this weakly coupled formulation, we then use holomorphy to match this to the
one loop approximation of the current correlator of the original system.

\subsubsection{Review of $\mathcal{N}=2$ Correlators}

To illustrate the general procedure, let us first review the computation of
current correlators for the $\mathcal{N}=2$ rank 1 $E_{8}$ SCFT \cite{GanorKrogh}
(see also \cite{Argyres:2007cn,Aharony:2007dj,Argyres:2007tq}).
Starting from the $E_{8}$ singularity:%
\begin{equation}
y^{2}=x^{3}+z^{5},
\end{equation}
we consider a complex deformation of this singularity:%
\begin{equation}
y^{2}=x^{3}+z^{5}+\left(  \delta f\right)  x+\left(  \delta g\right)  ,
\end{equation}
such that the seven-brane gauge theory of the deformed geometry, and its
coupling to the D3-brane probe admits a weakly
coupled description. The precise type of deformation is immaterial, so to
illustrate the general idea we consider a deformation of $E_{8}$ down to
$SU(5)$.\footnote{In \cite{GanorKrogh} the case of a deformation
to $SO(8)$ as well as $SO(10)$ was treated. The example we present is a
straightforward extension of this analysis.} The probe
D3-brane then corresponds to introducing a vector-like pair in the
$5\oplus\overline{5}$ into the $SU(5)$ gauge theory. \ Separating the D3-brane
off of the seven-brane gives a mass to this vector-like pair. The mass is
controlled by the value of the Coulomb branch parameter for the D3-brane,
$\widetilde{z}$. The one-loop running of the $SU(5)$ gauge coupling constant
as a function of $\widetilde{z}$ is then:%
\begin{equation}
\frac{4\pi}{g^{2}\left(  \widetilde{z}\right)  }=\frac{4\pi}{g^{2}\left(
\widetilde{\Lambda}\right)  }+\frac{1}{2\pi}\log\left(  \frac{\widetilde{\Lambda}%
}{\widetilde{z}}\right)  \text{.} \label{runner}%
\end{equation}
We now match this low energy behavior to the original $E_{8}$ gauge theory.
The main point is that holomorphy relates the Coulomb branch parameter
$\widetilde{z}$ of the low energy theory to $z$, the Coulomb branch parameter
of the high energy theory by:%
\begin{equation}
M^{\Delta-1}\widetilde{z}=z
\end{equation}
where $\Delta$ is the dimension of $z$. In other words, the running of the
gauge coupling constant as a function of $z$ is:%
\begin{equation}
\frac{4\pi}{g^{2}\left(  z\right)  }=\frac{4\pi}{g^{2}\left(  \widetilde{\Lambda}\right)
}+\frac{1}{2\pi}\log\left(  \frac{\widetilde{\Lambda}M^{\Delta-1}}{z}\right)
\text{.}%
\end{equation}
To relate this to the scales of the probe theory, we now use the fact
that $z$ has scaling dimension $\Delta$. In other words, $z^{1/\Delta}$
corresponds to a mass scale. This means that upon setting $\mu=z^{1/\Delta}$,
we learn:%
\begin{equation}
\frac{4\pi}{g^{2}\left(  \mu\right)  }=\frac{4\pi}{g(\Lambda)^{2}}+\frac{\Delta
}{2\pi}\log\left(  \frac{\Lambda}{\mu}\right)  \text{,}%
\end{equation}
where $\Lambda=\left(  \widetilde{\Lambda}M^{\Delta-1}\right)  ^{1/\Delta}$.
In other words, the coefficient $\rho$ of equation (\ref{currentcorrelator})
is:%
\begin{equation}
\rho= - \frac{\Delta}{8\pi^{2}}\text{.}%
\end{equation}
For the specific case of the $E_{8}$ SCFT, we have \cite{GanorKrogh} (see also
\cite{Aharony:2007dj,Argyres:2007cn,Argyres:2007tq}):%
\begin{equation}
\rho= - \frac{6}{8\pi^{2}}= - \frac{3}{4\pi^{2}}\text{.}%
\end{equation}
In other words, it as if we have $6$ vector-like pairs of $5\oplus\overline
{5}$ contributing in the $SU(5)$ theory.

From the perspective of the $E_8$ gauge theory, this is a quite striking result. Indeed, though the
contribution to the beta function of an $SU(5)$ gauge theory looks like an integral number of particles,
in the original $E_8$ gauge theory, the contribution of a hypermultiplet in the $248$ of $E_8$ is:
\begin{equation}
\rho_{\rm{hyper}} = - \frac{C_{2}(E_{8})}{4 \pi^{2}} = - \frac{30}{4 \pi^{2}},
\end{equation}
In other words, the actual contribution to the $E_8$ beta function is $1/10$ of a fundamental hypermultiplet!
Taking the $E_8$ gauge theory as a toy Standard Model, we see that at sufficiently
high energies, the gauge fields would appear to couple to an effectively
non-integer number of particles.

\subsubsection{$\mathcal{N}=1$ Correlators}

In this subsection we consider the analogous computation of beta functions for $\mathcal{N} = 1$ probe theories
of a Yukawa point. The main point is that in the above
computation of $\mathcal{N} = 2$ current correlators, we only
relied on general holomorphy considerations. In other
words, once we determine the scaling of $\widetilde{z}$ in the low energy theory,
matching to the parameter $z$ then specifies the holomorphic gauge coupling.

To determine the overall $\widetilde{z}$ scaling in the current correlator,
let us return to the starting configuration given by a D3-brane sitting at a
Yukawa point of an $SU(5)$ seven-brane. Moving the D3-brane parallel to the
seven-brane, but off of the Yukawa point, the contribution from the probe to
the beta function of $SU(5)$ is given by a massless vector-like pair in the
$5\oplus\overline{5}$. Moving the D3-brane off of the seven-brane gives a
mass to this vector-like pair, and affects the running of the $SU(5)$ gauge
theory, just as in equation (\ref{runner}). Performing a match between the
Coulomb branch parameter $\widetilde{z}$ to the Coulomb branch parameter of
the CFT point, we obtain the current correlator of the $\mathcal{N}=1$ CFT. In
other words, we learn that the contribution to the running of the couplings
from the probe sitting at the Yukawa point is:%
\begin{equation}
\frac{4\pi}{g^{2}\left(  \mu\right)  }=\frac{4\pi}{g(\Lambda)^{2}}+\frac{\Delta
}{2\pi}\log\left(  \frac{\Lambda}{\mu}\right)  \text{,} \label{n1runner}%
\end{equation}
where $\Delta$ is now the scaling dimension of the Coulomb branch parameter
$z$ of the $\mathcal{N} = 1$ theory.

An alternative argument for realizing the same scaling behavior is as follows.
Let us return to the configuration given by a D3-brane sitting at the Yukawa
point of the seven-brane configuration. At generic points of the seven-brane,
this corresponds to an $A_{4}$ singularity fibered over $z=0$. Now consider a complex deformation of the
Weierstrass model to an $A_{3}$ singularity. For a generic deformation with no
$z_{1}$ and $z_{2}$ dependence, this deformation also eliminates the presence
of the matter curves. Computing the contribution of the probe to the running
of the $SU(4)$ gauge theory, we find a vector-like pair in the $4 \oplus \overline
{4}$. Matching to the undeformed theory, we again conclude that the
$\widetilde{z}$ and $z$ dependence is as before, and we recover equation
(\ref{n1runner}).

From the above analysis, we see that the effective threshold is determined by
the scaling dimension of $z$, though the specific size of the threshold
correction depends on details of the probe theory. In particular, since we do not expect $z$ to have a
very high scaling dimension, the contribution from the probe theory does not induce a Landau pole,
and in particular preserves perturbative gauge coupling unification.\footnote{In principle,
there is a logical possibility that over the small amount of running between $M_{\cancel{CFT}}$
and the scale where conformal symmetry is restored, there is a sizable threshold correction which shuts off once we
enter the CFT regime. Though we cannot exclude this possibility, it seems rather implausible as the putative
large threshold would have to quickly turn on and then almost immediately switch off as a function of energy scales.}
For example, in the case of trivial monodromy, we have that $\Delta = 3$ for an $E_6$-point probe theory,
while $\Delta = 6$ for an $E_8$-point probe theory, which contribute as much as respectively one and two vector-like
pairs in the $10 \oplus \overline{10}$. Our expectation is that the presence of higher order monodromy will lead to a smaller
contribution to the running of the couplings. To fully address this question requires a more detailed analysis
of the resulting CFTs \cite{HTVW}.

\subsection{Coupling to the ``Hidden'' $U(1)_{D3}$}

As we have already mentioned, the $3-7$ strings must have mass of at least a few hundred GeV in order to avoid
conflict with experiment. A natural way to achieve this is to move onto the Coulomb branch of the probe theory so that
$z \neq 0$. In so doing, the $3-7$ strings will now develop a mass on the order of:
\begin{equation}
M_{3-7} \sim z^{1/\Delta}
\end{equation}
with $\Delta$ the scaling dimension of $z$. One can envision that this mass scale
is generically of the GUT scale, though it could also be much lower, if it is
set by supersymmetry breaking effects.

Moving onto the Coulomb branch leaves us with a $U(1)_{D3}$ gauge theory
with a tower of electric and magnetic states charged under $U(1)_{D3}$, as well as the gauge group
$SU(5)_{GUT} \times G_{\rm{extra}}$, where $G_{\rm{extra}}$ denotes the gauge group preserved by the
unfolding of the geometry. For example, in many cases it is desirable for phenomenological purposes
to retain a $U(1)$ Peccei-Quinn sector as well.

Based on general considerations, we expect there to be some amount of kinetic mixing between
this $U(1)_{D3}$ gauge boson and $U(1)_{Y}$. One might at first think that such mixing
is forbidden because $U(1)_{Y}$ is embedded inside of $SU(5)_{GUT}$. This first expectation is
incorrect because we can consider higher dimension operators which couple
the field strengths of $U(1)_{Y}$ and $U(1)_{D3}$ to the GUT breaking fluxes of an F-theory GUT \cite{EPOINT} (see 
also \cite{Goodsell:2009xc}). Thus, we generically expect kinetic mixing 
terms of the form \cite{Holdom:1985ag,Babu:1996vt,Dienes:1996zr,Babu:1997st}:
\begin{equation}
\delta L_{kin} = \int d^{2} \theta \,\, \varepsilon \mathcal{W}^{\alpha}_{Y}\mathcal{W}^{D3}_{\alpha} + h.c.
\end{equation}
where in a holomorphic basis of fields, $\varepsilon$ has logarithmic dependence
on the mass of the $3-7$ strings charged under both $U(1)_{Y}$ and $U(1)_{D3}$. Such kinetic mixing 
terms have been considered in the context of string based models, as for example in 
\cite{Dienes:1996zr,Lust:2003ky,Abel:2003ue,Abel:2004rp,Abel:2006qt,Abel:2008ai}.

To have evaded detection, it is necessary for this $U(1)_{D3}$ gauge boson to have a mass.
The mass of this $U(1)_{D3}$ gauge boson is in turn specified by the
vevs of $3-7$ strings stretched between the D3-brane, and either the GUT
brane stack, or the other $G_{\rm{extra}}$ seven-branes. In the former case, the $3-7$ strings
might also participate in electroweak symmetry breaking, though it is not as clear in this case
how to also generate masses for the Standard Model particles. The more innocuous possibility is that a $3-7$ string
which attaches to the $G_{\rm{extra}}$ seven-brane develops a vev, which can give a mass to $U(1)_{D3}$.

A novel feature of the kinetic mixing in the present system is that
because the $U(1)_{D3}$ gauge theory is strongly coupled,
we can generically expect both electric and magnetic states
to participate in kinetic mixing. Including for this possibility, we see that
it is natural to expect there to be kinetic mixing between both the electric and magnetic
dual field strengths:
\begin{equation}
L_{eff} \supset \varepsilon_{\rm{elec}} F^{Y}_{\mu \nu} F_{D3}^{\mu \nu} + \varepsilon_{\rm{mag}} F^{Y}_{\mu \nu} \widetilde{F}^{\mu \nu}_{D3}.
\end{equation}
It would be interesting to study the phenomenological
consequences of this sort of effect in more detail
\cite{WorkProgHV}.\footnote{As far as we are aware, the phenomenology
of magnetic kinetic mixing is a recent possibility mentioned
for example in \cite{Brummer:2009cs,Bruemmer:2009ky,Benakli:2009mk} (see also \cite{WeinerTalk}
in the context of dark matter phenomenology).}

\section{Conclusions}\label{sec:CONCLUSIONS}

Though often viewed as a secondary ingredient in constructing the visible sector of an F-theory GUT,
D3-branes are often a necessary component of a global compactification. In this paper
we have seen that much of the structure already required for viable F-theory GUTs
also naturally suggests including D3-branes as an additional sector which is attracted
to the Yukawa points of the geometry. Utilizing the dictionary between the background
geometry and the worldvolume theory on a D3-brane, we have investigated the probe theories of
D3-branes sitting at E-type points. In addition, we have
studied how this probe theory couples to the Standard Model, both through F-terms, as well
as its coupling to the gauge fields of the Standard Model. In the remainder of
this section we discuss some further possible avenues of investigation.

To fully specify the way that the D3-brane probe couples to the Standard Model, it is necessary to extract additional information
about the low energy theory induced by more general $\mathcal{N} = 1$ deformations of the original $\mathcal{N} = 2$ probe
theories. Such deformations appear to be interesting both from the perspective of F-theory considerations, as well as from the purely
field theoretic perspective. The study of such CFTs will be presented elsewhere \cite{HTVW}.

Though we have focussed on the probe of a single D3-brane, more generally, one might consider
the probe theory of a large $N$ number of D3-branes. In particular,
it would be interesting to develop a precise holographic dual for such configurations as a further
tool to study such worldvolume theories. Though a general $\mathcal{N} = 1$ deformation may appear to lead
to a complicated supergravity dual, in at least one case, corresponding to trivial seven-brane monodromy,
we expect that the theory flows back to the original $\mathcal{N} = 2$ theory, with corresponding
holographic dual as in \cite{Aharony:1998xz}.

At the level of model building applications, coupling the Standard Model to the CFT has been considered
for various applications, both as a sector of interest in its own right, or as providing a set of
ingredients for potentially solving problems in both supersymmetric and non-supersymmetric model
building. As some possible examples, the presence of vector-like states suggests a potentially novel way to realize a
gauge mediation sector with the messengers in a strongly coupled CFT, along the lines suggested in \cite{GMGM}.
It would also be interesting to see if there is a natural mechanism to break supersymmetry on the D3-brane
probe and communicate it to the visible sector through gauge mediation.

Given that the motion of D3-branes in a compactification provides a natural set of inflaton candidates, it would be
interesting to see if our brane probe can also play such a role. Along these lines, the mode describing motion normal to the D3-brane could play the role of the inflaton, and inflation would end as the D3-brane reaches the Yukawa point, reheating through its couplings to the visible sector.
As a related possibility, the worldvolume theory of an anti-D3-brane may also provide a starting point for realizing
an inflationary scenario. Indeed, for an isolated seven-brane, an anti-D3-brane probe is still supersymmetric, though it
preserves a different set of supercharges from a probe D3-brane. Tilting the seven-branes to an $\mathcal{N} = 1$ configuration
would break supersymmetry. In this case, inflation would end after the anti-D3-brane dissolves into the Higgs branch of the seven-brane,
decreasing the instanton number on the seven-brane by one unit.

\section*{Acknowledgements}

We thank Y. Tachikawa and B. Wecht for many helpful discussions, and collaboration on
related work. We also thank C. C\'{o}rdova, D. Green, K. Intriligator, S-J. Rey, N. Seiberg, M.J. Strassler
and M. Wijnholt for helpful discussions. JJH thanks the Harvard high energy theory group
for generous hospitality during part of this work. The work of JJH is supported by
NSF grant PHY-0503584. The work of CV is supported by
NSF grant PHY-0244281.

\newpage
\bibliographystyle{ssg}
\bibliography{funparticles}

\end{document}